**Effect of the Lattice-distortion on the Electronic Structure, Magnetic Anisotropy, and Hall Conductivities of the CoFeCrGa Spin Gapless Semiconductor: A First-Principles Study**

Amar Kumar[1], Sujeet Chaudhary[1*], and Sharat Chandra[2*]

[1]Thin Film Laboratory, Department of Physics, Indian Institute of Technology Delhi, New Delhi 110016, India

[2]Material Science Group, Indira Gandhi Centre for Atomic Research, a *CI* of Homi Bhabha National Institute (Mumbai), Kalpakkam, Tamil Nadu-603102, India

*Corresponding Authors: sujeetc@physics.iitd.ac.in and sharat@igcar.gov.in

**Abstract**

Spin gapless semiconductors (SGSs), novel quantum materials, are notable for their tunable spin-transport properties. Considering that the SGS materials might have an invariably deformed lattice upon integration into devices, and given that the SGS nature is highly sensitive to external factors, the impact of lattice distortions on the different physical properties of CoFeCrGa SGS alloy has been investigated using density functional theory calculations. For lattice distortions, the uniform strain corresponding to $-6\% \leq \Delta V/V_0 \leq 6\%$ (*a*: 5.60–5.83 Å), and the tetragonal distortion corresponding to $0.8 \leq c/a \leq 1.2$ (*a*: 5.38–6.16 Å, *c*: 4.92–6.45 Å) are modelled. All uniformly strained CoFeCrGa structures are found to display SGS character, magnetic isotropy, small anomalous Hall conductivity (AHC), and small spin Hall conductivity (SHC) - closely resembling those of the ideal CoFeCrGa structure. In contrast, the tetragonally deformed structures display nearly half-metallic behavior with very high spin polarization, very large magnetic anisotropy (~ $10^6$ J/m$^3$), and very large AHC ranging from -215 to 250 S/cm - depending on the axial ratio of the distorted structure. The SHC, however, does not change significantly under tetragonal distortion and remains nearly of the same order as that of the Y-I ordered structure. In summary, these findings demonstrate that CoFeCrGa displays favorable spintronic properties even under lattice distortions, underscoring its potential for next-generation spintronic applications.

**Effect of the Lattice-distortion on the Electronic Structure, Magnetic Anisotropy, and Hall Conductivities of the CoFeCrGa Spin Gapless Semiconductor: A First-Principles Study**

Amar Kumar[1], Sujeet Chaudhary[1*], and Sharat Chandra[2*]

[1]Thin Film Laboratory, Department of Physics, Indian Institute of Technology Delhi, New Delhi 110016, India

[2]Material Science Group, Indira Gandhi Centre for Atomic Research, a *CI* of Homi Bhabha National Institute (Mumbai), Kalpakkam, Tamil Nadu-603102, India

*Corresponding Authors: sujeetc@physics.iitd.ac.in and sharat@igcar.gov.in

## 1. Introduction

The prediction and design of novel magnetic materials with favorable spintronic properties, such as - high spin polarization ($P$), high Curie temperature ($T_C$), large magnetic anisotropy (MA), and large Hall conductivities - are crucial for enabling advanced spintronic applications. Consequently, over the past several decades, numerous classes for magnetic materials have been proposed and extensively studied to fulfill criteria for spintronic applications. The most promising spintronics material classes to date include (but are not limited to) - alloys of Co, Fe, and B; half-metallic perovskite materials; diluted magnetic semiconductors, and half-metallic cubic and tetragonal Heusler alloys (HAs) [1]. However, each material class exhibits inherent drawbacks upon device integration, which limit the efficiency of spintronic devices. For example, magnetic tunnel junctions employing CoFeB alloys as ferromagnetic (FM) material layer suffer from weak interfacial perpendicular magnetic anisotropy (PMA) at the interfaces with the tunnel barrier layers [2], whereas those using tetragonal Heusler alloys display low tunnelling magnetoresistance [3–5]. On the other hand, perovskite materials often exhibited structural instabilities in the thin film form, magnetic semiconductors have small Curie temperatures, and cubic HAs display low experimental spin polarization along with magnetic isotropic nature [6–8]. Such limitations hinder the overall performance of spintronic devices. While the search for the solution to these existing problems is still ongoing, these limitations have also motivated material scientists to explore new conceptual materials with favorable spintronic properties.

In this context, spin gapless semiconductors (SGS) represent a recently discovered material category belonging to the Heusler alloy family. Unlike conventional half-metallic HAs, SGS materials exhibit a unique electronic band structure characterized by a zero (closed) gap in one spin channel and a finite bandgap (similar to semiconductors) in the other spin channel. In terms of material properties, the main experimental signatures of SGSs include small and nearly temperature-independent anomalous Hall conductivities, linear magnetoresistance at low field and temperature, and nearly temperature-independent charge carrier conductivity. Owing to their distinctive unique band structure, SGS offer several advantages: (i) higher $T_C$ compared to the magnetic semiconductors, (ii) simultaneous availability of electrons as well as holes as 100% spin-polarized carriers depending on their band structure, (iii) energy-efficient transport due to low excitation energies, (iv) efficient spin transport arising from long diffusion lengths of charge carriers, and (v) efficient spin injection into semiconductors due to minimum conductivity mismatch. Thus, by combining the properties of perfect half-metallic ferromagnets (HMFs) and semiconductors (SCs), SGS serves as a bridge between HMFs and SCs, making them highly promising for tunable spin transport applications in next-generation spintronic applications [9,10].

Moreover, due to their distinct band structure, the physical properties of SGS materials are reported to be highly susceptible to external perturbations such as electric and magnetic fields, structural imperfections, temperature, stress, *etc*. Among these, stress and temperature stand out as readily accessible factors during the experimental growth process, typically encountered through common mechanisms such as lattice mismatch with the substrate or adjacent layer(s), specific deposition technique employed, thermal treatment of the deposited film, *etc*. These factors often lead to distorted lattices, which in turn, can modify the structural, electronic, and magnetic properties of the SGS material, sometimes even leading to entirely different physical property profiles. Interestingly, lattice distortions can also induce useful phenomena, such as magnetic anisotropy, Hall conductivities, magnetoresistance effects, as reported in several cases. Additionally, under the lattice distortion, the material may have stable or metastable tetragonally distorted phases, which could appear in experiments in the form of single or mixed phases. Thus, lattice distortions may enable the experimental realization of favorable spintronic properties. Therefore, studying the impact of lattice distortions on the physical properties of SGS materials is not only necessary but also seems to be beneficial for device design and warrants a thorough investigation [7,11–16].

Over the past decade, numerous SGSs have been predicted through first-principle calculations, and among them, several have been experimentally demonstrated. Most of the experimentally fabricated SGS belong to the HAs family, for example – $Cr_3Al$, $V_3Al$, FeMnGa/Al/In, $Mn_2CoAl$, $Ti_2CoSi$, $Ti_2MnAl$, $Ti_2CrSi$/Sn, CoFeMnSi, CoFeCrGa, CoFeCrAl, *etc* [1,17]. Among these also, SGS materials belonging to the quaternary Heusler alloy family hold greater potential for spintronics than SGSs in binary or ternary Heusler alloys, owing to their higher ability to display tunable electronic and magnetic properties [10]. In this context, CoFeCrGa is a recently discovered SGS belonging to the quaternary HAs family, which was first theoretically predicted and later experimentally fabricated, exhibiting a high $T_C > 600$ K and $M_S$ ~$2.0\mu_B$/*f.u.*, along with SGS nature [18,19]. Another noteworthy and unexpected aspect of CoFeCrGa is that it preserves the SGS nature even in thin film form, a characteristic often destroyed for most materials in thin films [20]. However, despite being such an important material, CoFeCrGa is relatively underexplored, with few existing studies primarily focusing on probing the electronic nature of bulk structure and thin films, and magneto-transport properties like anomalous Hall conductivities and magnetoresistance [18–20]. Regarding the lattice distortion in CoFeCrGa, Bainsla *et al.* have studied the electronic band structure and magnetization of CoFeCrGa at different lattice parameters (*a*) – 5.40 Å, 5.50 Å, 5.60 Å, 5.71 Å, and 5.79 Å – using density functional theory (DFT) calculations. Their study, however, provides a preliminary overview, showing that the SGS nature is lost as '*a*' decreases from the experimental value of 5.79 Å, while the magnetization remains nearly constant across the considered lattice parameters [19]. However, the literature still lacks a comprehensive and systematic investigation of how lattice distortions affect the different physical properties of CoFeCrGa, which is crucial for its potential spintronic applications.

Therefore, in the present study, we have tried to examine the impact of lattice distortion on various electronic and magnetic properties of CoFeCrGa, under two kinds of lattice distortions -uniform strain and tetragonal distortions (uniaxial strain along *c*-axis and uniform bi-axial strain in the *a-b* plane). We also explore the possibility of stable or metastable tetragonal phase for the lattice distorted CoFeCrGa structures, with the help of state-of-the-art DFT calculations. Both the uniform strain and tetragonal distortion have been modelled within the thermodynamically feasible range for experimental relevance, as discussed at the beginning of the relevant sub-sections in Section 3. The rest of this article is organized as follows: First, the computational methodology employed for the present work is briefly discussed in Section 2. Then, the validation of the exchange-correlation functional chosen for studying the physical properties of the CoFeCrGa alloy is discussed in Section 3.1. Subsequently, Section 3.2 presents the results for the impact of uniform strains and tetragonal distortions on the electronic structure and magnetization of CoFeCrGa. Also, the consequences of this distortion - magnetic anisotropies and Hall conductivities - have been discussed in detail to evaluate the spintronic potential of CoFeCrGa. Finally, the concluding section, Section 4, comprehensively summarizes all the findings.

## 2. Computational methodology

The plane-wave pseudopotential-based DFT calculations are performed using the QUANTUM ESPRESSO package to gain insight into the effect of lattice distortion on the structural, electronic, and magnetic properties of the CoFeCrGa alloy [21,22]. Generalized gradient approximation (GGA) and GGA+U methods were initially employed to validate the choice of exchange-correlation functional, with $U$ as the Hubbard parameter, however, all final calculations reported in this work are performed using GGA, as described in subsection 3.1 [23,24]. For the *Y-I* ordered 16-atom cubic unit-cell, the plane wave kinetic energy cut-off of 350 Ry and a 15×15×15 $k$-point mesh are used, along with the optimized tetrahedron method for the Brillouin zone integration [25]. For varied-size unit cells, whenever used, the $k$-points mesh is adjusted accordingly to maintain consistent k-point density to ensure comparable outcomes. For approximating the atomic potential, the pseudopotentials from PSlibrary are used, with valence electronic configuration of Co ($3s^2 4s^2 3p^6 3d^7$), Fe ($3s^2 4s^2 3p^6 3d^6$), Cr ($3s^2 4s^2 3p^6 3d^4$), and Ga ($4s^2 3d^{10} 4p^1$) [26]. To achieve the minimum energy configurations for the utilized crystal structures, atomic relaxations are performed using the Davidson iterative diagonalization method until the total force falls below $10^{-3}$ Ry/Bohr; while the self-consistency for the electron density is achieved when change in total energy between the successive cycles are less than $10^{-6}$ Ry. Transverse Hall phenomena, including the anomalous and spin Hall effects, are investigated using a tight-binding approach based on maximally localized Wannier functions (MLWFs), as implemented in the Wannier90 package [27].

For the calculation of structural properties, electronic structure, and magnetization (subsection 3.2), only collinear spins are considered for computational simplicity, as the inclusion of spin-orbit coupling (SOC) has been reported to have a minor effect on these properties for similar materials, as indicated in the previous studies [28–30]. Whereas, for the calculation of magnetic anisotropies and Hall conductivities, SOC is explicitly included through the fully relativistic pseudopotentials, as it is a primary contributor to these properties. Further computational details regarding the calculation of magnetic anisotropies and Hall conductivities are provided in subsections 3.3 and 3.4.

## 3. Result and Discussion

### 3.1 Exchange Correlation functional validation for CoFeCrGa

As widely known, there are various approximations for exchange-correlation (XC) functionals, and the accuracy of DFT outcomes strongly depends on the chosen XC functional for calculations. The GGA, within the parameterization of the Perdew–Burke–Ernzerhof (PBE), provides reasonably accurate results at a low computational cost, making it the most commonly employed XC for metallic and semiconducting materials. Within GGA, the electronic interactions are treated in an average manner by means of mean-field approximation. However, since CoFeCrGa contains 3*d*-transition metals (i.e., Co, Fe, and Cr), accounting for the strong on-site Coulomb interactions among the localized *d*-electrons is widely recommended in the literature, as it can sometimes yield a more accurate description of the material's physical properties. This is typically achieved by integrating the GGA with Hubbard parameter ($U$), which represents the on-site Coulomb correction for the *d*-orbitals of the transition metals. Accordingly, we have calculated the various ground state physical properties of CoFeCrGa with DFT+U method [31]. For CoFeCrGa, the calculated U parameters for the Co, Fe, and Cr atoms in *Y-I* ordered structure are: $U_{\mathrm{Co}}$ = 6.53 eV, $U_{\mathrm{Fe}}$ = 5.28 eV, and $U_{\mathrm{Cr}}$ = 5.70 eV. At once, these $U$ values may appear relatively large, when compared to the reported range of 2.5 eV – 3.5 eV for 3*d*-transition metals. However, it is noticeable that the magnitude of $U$ values depends on several factors, such as atomic potentials, basis-set, calculation method and parameters, *etc*. Therefore, U values are not fixed constants and can have different values with different inputs for the same element [32]. Then, using the GGA and GGA+U methods, different fundamental physical properties are calculated and compared them with the available experimental and theoretical findings, to determine a suitable XC for studying the physical properties of CoFeCrGa.

**Table 1:** The calculated Hubbard parameters ($U_{\mathrm{Co}}$ for Co atoms, $U_{\mathrm{Fe}}$ for Fe atoms, and $U_{\mathrm{Cr}}$ for Cr atoms), optimized lattice parameter $a$ (Å), total magnetic moment $M_s$ ($\mu_B$/*f.u.*), atomic magnetic moments ($m_{Co}$, $m_{Fe}$, $m_{Cr}$, $m_{Ga}$ for Co, Fe, Cr and Ga, respectively), SGS signature, and spin polarization ($P$) for *Y-I*-ordered CoFeCrGa with GGA and GGA+U methods. The other theoretical and experimental results are also provided for comparison. Here, 'Exp.' denotes experimental data, and the lattice parameters marked with an asterisk (*) indicate that the fixed experimental lattice parameter is used for calculating the electronic and magnetic properties.

| **Properties** | **Present work (GGA)** | **Present work (GGA+U)** | **Present work (GGA+U)** | **Shi *et al.* [33] (GGA)** | **Bainsla *et al.* [19]** | | **Mishra *et al.* [20] (Exp.)** |
|---|---|---|---|---|---|---|---|
| | | | | | **GGA** | **Exp.** | |
| **$U_{\mathrm{Co}}$(eV)** | - | 6.53 | 1.92 | - | - | - | - |
| **$U_{\mathrm{Fe}}$(eV)** | - | 5.28 | 1.80 | - | - | - | - |
| **$U_{\mathrm{Cr}}$(eV)** | - | 5.70 | 1.59 | - | - | - | - |
| ***a* (Å)** | 5.72 | 6.48 | 5.76 | 5.72 | 5.79* | 5.79* | 5.72 |
| **$M_s$ ($\mu_B$/*f.u.*)** | 2.00 | 9.33 | 2.65 | 1.97 | 1.98 | 2.01 | 1.86 |
| **$m_{Co}$($\mu_B$)** | 0.99 | 1.67 | 0.92, 0.96 | 0.94 | - | - | - |
| **$m_{Fe}$($\mu_B$)** | -0.70 | 3.32 | 1.79, 2.06 | −0.74 | - | - | - |
| **$m_{Cr}$($\mu_B$)** | 1.68 | 3.86 | 0.07 | 1.81 | - | - | - |
| **$m_{Ga}$($\mu_B$)** | -0.03 | -0.11 | -0.05 | −0.04 | - | - | - |
| ***SGS signature*** | Yes | No | No | Yes | Yes | Yes | Yes |
| ***Polarization (P)*** | 92.33% | 33.14% | 23.89% | ~100% | - | - | - |

As seen from Table 1, when employing the GGA XC functional, CoFeCrGa crystallized in the *Y-I* type structure with an optimized lattice parameter of 5.72 Å, a total magnetic moment of 2.0 $\mu_B$/*f.u.* along with SGS nature. In the Y-I ordered cubic structure, Fe, Cr, Co, and Ga occupy the 4a (0, 0, 0), 4b (0.25, 0.25, 0.25), 4c (0.50, 0.50, 0.50), and 4d (0.75, 0.75, 0.75) Wyckoff positions, respectively, in the $F\overline{4}3m$ (No. 216) space group. Notably, all these findings, using GGA, are closer to the experimental results, as illustrated from Table 1 [20]. Whereas the calculated properties using the GGA+U method significantly differ from the experimental and GGA results. Therefore, GGA looks like a better option for investigating the physical properties of CoFeCrGa alloy. For further affirmation of the better XC functional, the above-mentioned properties were also calculated using another *U* values from Gao *et al.*, which range the *U* parameters for all atoms between 1.5 eV - 2.0 eV [18]. However, even with these *U* values, the physical properties of CoFeCrGa differ significantly from the findings of the experimental and GGA methods. This leads to the conclusion that the GGA+U method *with the specified U values* is not beneficial for studying the physical properties of CoFeCrGa alloy. On the other hand, GGA provides results consistent with the experiments, and seems to be an adequate XC for the calculations, and will therefore be used for further calculations. Notably, similar observations have been reported for many Co-based Heusler alloys, where GGA is found to be more suitable than GGA+U, for studying their physical properties [34–38]. Thus, the present findings appear reasonable. Thereby, GGA will be utilized for all further calculations in the present study.

### 3.2 Effect of lattice distortions on structural, electronic, and magnetic properties of CoFeCrGa

The following section discusses the impact of lattice distortions on the structural, electronic, and magnetic properties of CoFeCrGa alloy. Both uniformly strained and tetragonally distorted structures are modelled using a 16 atoms unit-cell. The optimized unit-cell volume of the *Y-I* ordered structure ($V_0$) is uniformly contracted and expanded from ($V_0$ - 6%$V_0$) to ($V_0$ + 6%$V_0$) in a step size of 1%$V_0$ to model the uniformly (or isotopically) strained structures. Here $V_0 = (a_0)^3$; where $a_0$ = 5.72 Å is the optimized lattice parameter for the *Y-I* ordered CoFeCrGa alloy [20]. The uniform strain can be defined as a percentage change of the optimized volume, *i.e.,* % $\Delta V/V_0$ with $\Delta V = (V-V_0)$. This uniform strain leads to lattice parameters ranging between 5.60Å - 5.83Å (Fig. 1(a)). Furthermore, the tetragonal (or uniaxial) distortions are modelled by setting cell parameters as $a = b$ and varying $0.8 \leq c/a \leq 1.2$, with fixed unit-cell volume at $V_0$. The tetragonal distortion can be defined in terms of $\Delta c/a$, *i.e.*, variation of axial ratio ($c/a$) from 1.00, if needed. This tetragonal distortion results in lattice parameters for the distorted unit-cell ranging between 5.38 Å $\leq a \leq$ 6.16 Å and 4.92 Å $\leq c \leq$ 6.45 Å (Fig. 3(a)).

The above-mentioned ranges of lattice parameter for the lattice distorted structures are considered, in light of various factors. As observed from many studies, the lattice mismatch of HAs with the adjacent layers, stress from growth techniques, and post-growth treatment generally lead to a uniform change in the lattice parameters of cubic HAs by 1% - 2%, which corresponds approximately to a volume change of ~5% - 6% relative to optimized volume. Therefore, a maximum change of ±6% in $V_0$ is considered for simulating the uniform strain. For modeling tetragonal distortions, the *c/a* value is made to vary between ~0.8-1.2, with the fixed $V_0$ unit-cell volume, in a step size of $|\Delta c/a| = 0.05$. Here, for modelling the tetragonal distortion, the distorted unit-cell volume is kept constant at $V_0$, because most of the full HAs exhibiting the stable- or metastable- tetragonal phase often maintain a volume similar to that of their global minimum cubic phase, or show only minor changes [14,39–41]. The following subsections will discuss the impact of uniform strain and tetragonal distortion on various physical properties of CoFeCrGa.

**3.2.1 Effect of uniform strain on structural, electronic, and magnetic properties of CoFeCrGa:**

The relative formation energy (RFE) of the lattice-distorted structures is an important parameter as it provides information about the likelihood of their occurrence, *i.e.,* the possibility of formation of during experimental growth. Additionally, it elucidates the relative stability of the distorted structures *w.r.t.* global ordered structure. Therefore, the relative formation energies of the lattice distorted structures, which sometimes also denoted as the distortion formation energy, have been computed *w.r.t.* the *Y-I* ordered structure using the following formula,

$$\mathrm{RFE(A)} = \mathrm{E(A)} - \mathrm{E(YI)} \tag{1}$$

Here, RFE(A), E(A), and E(YI) represent the RFE of the lattice distorted structure, the total energy of the lattice distorted structure (A), and the total energy of the optimized cubic structure (*Y-I* ordered), respectively. Figure 1 summarizes the structural, electronic, and magnetic properties of isotopically strained CoFeCrGa structures. As demonstrated in Figure 1(b), all uniformly strained structures exhibit positive RFE, which increases with increasing strain value ($\%|\Delta V/V_0|$); therefore, their growth is relatively less favorable than the optimized *Y-I* structures, and their occurrence probability decreases with increasing strain values. However, the uniformly strained structures within the range $-5\% \leq \Delta V/V_0 \leq 6\%$ exhibit very small RFE values ($\lesssim$ 0.1 eV/*f.u.*), which are comparable to the thermal energies typically encountered in experiments and also satisfy the distortion formation energy criterion for most of the experimentally observed strained full-Heusler alloys [42]. This inference suggests that the uniformly strained structures corresponding to $-5\% \leq \Delta V/V_0 \leq 6\%$ should readily form during the experimental growth of CoFeCrGa alloy. On the other hand, the strained structure with a larger strain value ($\Delta V/V_0 = -6\%$) has a slightly higher RFE, greater than 0.1 eV/*f.u.,* and therefore should have a smaller occurrence probability than the other strained structures.

Thereafter, the effect of strain on the electronic properties of CoFeCrGa is investigated using total and partial density of states (DOS). The spin-polarized DOS under uniform strain are shown in Figure 2, where we present DOS only for a few strain values, *viz* $\Delta V/V_0 = 0, \pm 2\%, \pm 4\%$, and $\pm 6\%$. Before describing the influence of the lattice distortions on the electronic properties, let us first discuss some key features of DOS for the ideal (*Y-I* ordered) structure of the CoFeCrGa alloy. Similar to the DOS plot for other Co-based Heusler compounds, the DOS plot of CoFeCrGa also exhibits the peak and valley characteristics resulting from the *d*-orbitals localization and van Hove singularities [13]. Referring to the DOS of CoFeCrGa from supplemental material Figures S1(a) and from our previous study [20], it is evident that *near* the Fermi level ($E_F$), the valence band maxima and conduction band minima nearly touch each other (or slightly overlapped) for spin-up electrons, forming a spin-zero gap point. Meanwhile, a finite band gap of ~125 meV exists for spin-down electrons. These features imply that CoFeCrGa has a spin-gapless semiconducting nature [20]. Here, it is worth noting that both of these band characteristics, *i.e.*, the location of the zero-gap point in the spin-up band and the finite band gap in the spin-down band, did not take place precisely at $E_F$ but occurred slightly away from $E_F$. Additionally, the up-spin DOS is also not exactly zero at the zero-gap location. More precisely, the minimum spin-up DOS occurs at 0.007 eV ($E$-$E_F$) with a value of 0.0645 states/eV-atom, and the minority spin-gap lies between 0.100 eV and 0.226 eV, relative to $E_F$. However, despite these slight deviations from the ideal SGS band structure (gapless point-location, DOS at gapless point, location of minority valence band edge), the electronic nature of CoFeCrGa can still be regarded as SGS, as the SGS nature of CoFeCrGa was also confirmed experimentally in our previous study through the transport measurements at the same lattice parameter. A similar scenario is also observed in ref. [19], where the SGS nature of CoFeCrGa was experimentally confirmed through transport measurements despite the presence of comparable theoretically calculated DOS at $E_F$ at the experimental lattice parameter. Both of these studies suggested that such a slight deviation in the band characteristics can be safely neglected, and this doesn't affect the decision about the SGS nature of CoFeCrGa. Thus, under these conditions, CoFeCrGa can be specified as SGS.

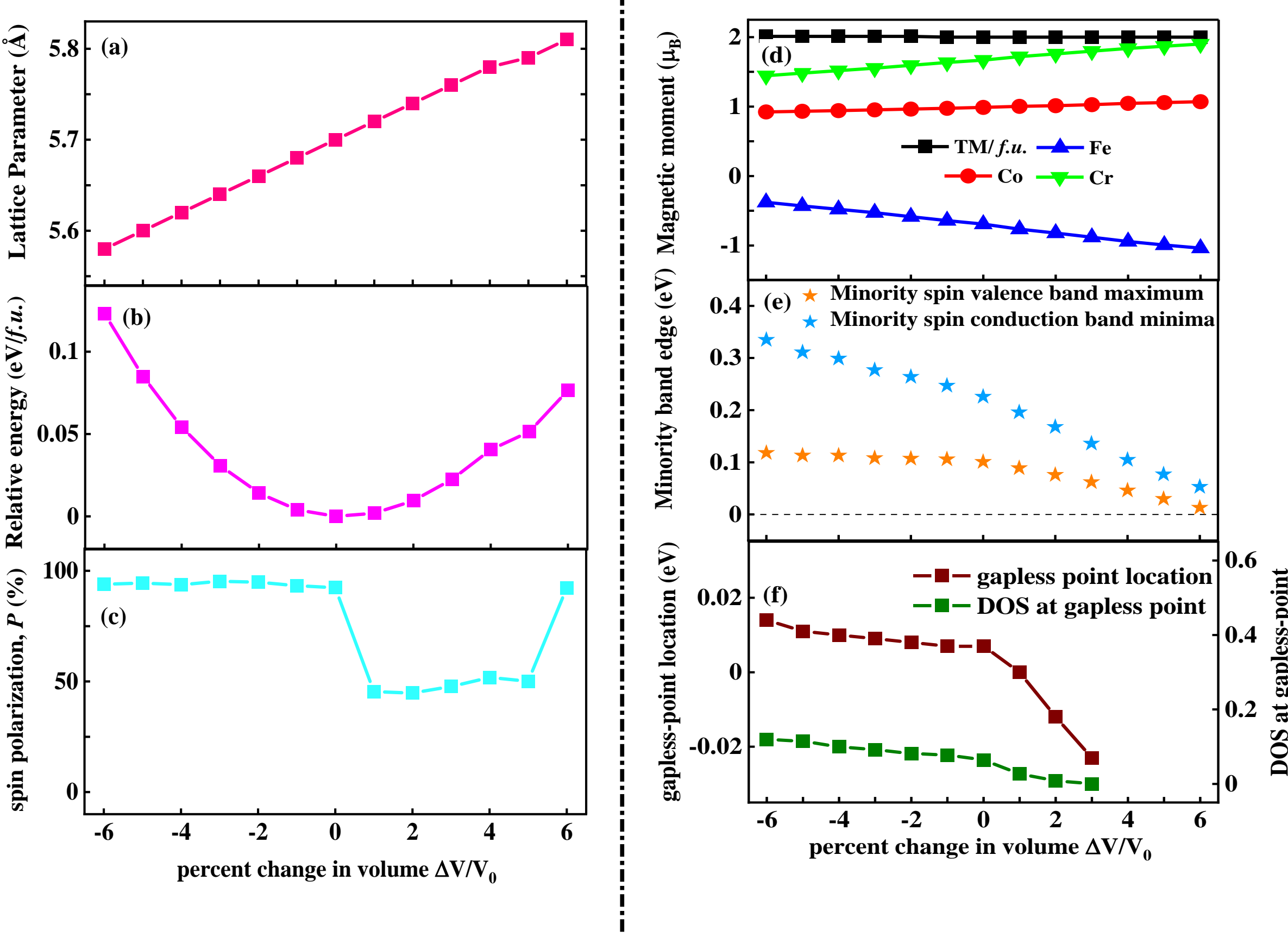

**Figure 1**: (a) Lattice parameter, (b) Relative formation energy, (c) spin polarization, (d) total and atomic magnetic moments, (e) minority valence band maxima and minority conduction band minima, and (f) gapless-point location *w.r.t.* Fermi level and DOS (***states/eV-atom***) at gapless-point for uniformly (or isotopically) strained CoFeCrGa alloy as a function of percent change in unit-cell volume (%$\Delta V/V_0$). The horizontal short-dashed line in Figure 1(f) indicates the Fermi level. See the text for more details.

Furthermore, similar to previous studies, certain relaxation criteria have been adopted in present study for defining the SGS characteristics when the calculated DOS deviates slightly from the ideal SGS profile, as a perfectly ideal SGS DOS is rarely realized in practice. In line with earlier reports, such acceptable deviations include: (i) a small DOS at the gapless point ($\leq$ 0.1 states/eV-atom), (ii) a slight shift of the spin zero-gap or gapless point relative to $E_F$ (< ±0.02 eV), (iii) a minor displacement of the minority valence band edge toward higher energies relative to $E_F$ ($\leq$ 0.1 eV), and (iv) a weak overlap between majority valence and conduction band edge at the gapless point ($\leq$ ±0.1 eV, where a negative value denotes a small gap at the gapless point). These considerations are consistent with earlier studies on other SGS alloys, and primarily arise from limited computational accuracy from the intrinsic approximations of the employed exchange–correlation functional and other input parameters used for calculations [19,20,43,44]. Therefore, under these conditions, the DOS can still be reliably classified as SGS. Furthermore, it is also noticeable that, in an ideal SGS, the spin polarization at $E_F$ is expected to be zero owing to the absence of DOS at $E_F$, as $E_F$ lies at the gapless point in the majority spin channel and within the bandgap in minority spin channel. However, due to a slight deviations from the ideal DOS, a small finite DOS is observed at $E_F$ in practice, leading to a nonzero spin polarization. Accordingly, spin polarization values have also been reported for different strained structures wherever relevant, as shown in Table 1 and Figure 1.

Now, let us discuss the impact of uniform strain on the electronic and magnetic properties of CoFeCrGa. Under uniform strain, CoFeCrGa retains an electronic structure (DOS profile) similar to that of the *Y-I* ordered structure, with only minor modifications in overall DOS shape and in states near $E_F$. From the perspective of the DOS shape, for negative strain (-6% $\leq \Delta V/V_0 \leq$ -1%), the reduced unit-cell volume of the strained structure leads to broader and shallower bands compared to the Y-I ordered structure, as illustrated in Figures 2(a)–(c). This broadening arises from the increased electronic orbitals' overlap due to the shorter interatomic distances in the strained structures. Conversely, the positive strain (1% $\leq \Delta V/V_0 \leq$ 6%) produces the opposite effect: decreased electronic orbitals' overlap from the expanded unit-cell volume results in narrower and peakier bands, as depicted in Figures 2(d) – (f). Nevertheless, the overall shape of the DOS across all strained structures remains largely unchanged, due to the similar crystallography as of the relaxed structures.

Regarding the states near $E_F$, only minor changes are observed in both spin channels. In the majority-spin channel, the position of the gapless point is slightly deviated from $E_F$, with finite DOS magnitude at the gapless point and overlapping of the majority valence and conduction bands. However, for all strained structures, these deviations from the ideal SGS DOS are small and remain within the relaxation criteria discussed earlier - the gapless point remains within ±0.02 eV of $E_F$, DOS at the gapless point stays very small ($\leq$ 0.1 states/eV-atom), and the majority valence-conduction band overlapping is limited to $\delta \leq 0.1$ eV. Thus, the majority spin band for all strained structures perseveres its gapless nature. Whereas, in the minority spin-channel, uniform strain affects the minority energy gap width ($E_{gap} = E_c - E_v$) and its center w.r.t. $E_F$. Here, $E_c$ and $E_v$ represent the valence band maxima and spin-down conduction band minima, in the minority spin channel. As seen from Figure 1(e), $E_{gap}$ decreases with increasing cell volume, reducing from 0.217 eV (-6%) to 0.040 eV (6%) at extreme strain values; and as shown in Figure 2, the minority band center shifts towards lower binding energies. These changes in minority spin DOS also lie within the relaxation limit. Therefore, despite these minor modifications, the overall electronic nature of CoFeCrGa remains unaffected, and all uniformly strained structures show the SGS nature. Importantly, all these subtle modifications in DOS shape and states around $E_F$ under uniform strain can be microscopically attributed to the changes in the exchange interactions and interatomic hybridizations with strain value, as explained for other Heusler alloys, like – $Co_2CrAl$, $Co_2FeAl$, and $Co_2MnGe$ in the literature [11,45,46].

Then the spin polarization at Fermi level is calculated through $P = (D1 - D2)/(D1 + D2)$, with $D1$ and $D2$ representing the majority and minority DOS at $E_F$. As there are modifications in DOS at the Fermi level, the spin polarization also changes accordingly for strained structures, since the $P$ is directly interrelated to the DOS at $E_F$. The calculated spin polarization, gapless point location, DOS at gapless point, minority valence band maxima, minority conduction band minima, atomic magnetic moments, and total magnetic moments for the uniformly strained structures are given in Figures 1(b) – 1(f). For the negative strains ($-6\% \leq \Delta V/V_0 < 0\%$), CoFeCrGa possesses ≈100% polarization. Above that, as the strain value changes between 1% - 5%, $P$ decreases to ~50%. For highest positive strain ($\Delta V/V_0 = 6\%$), $P$ again increases to ~100%.

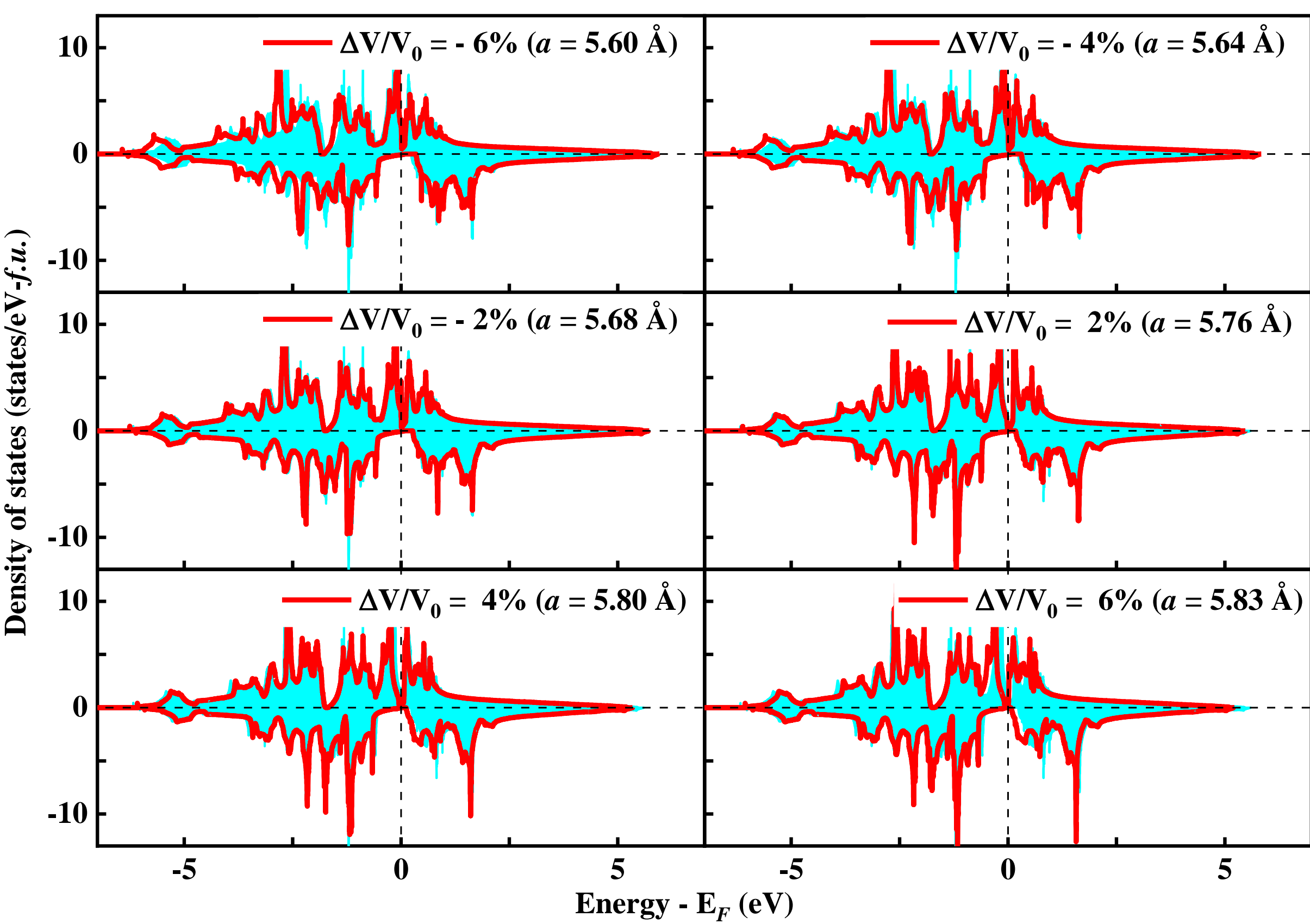

**Figure 2**: The density of states plots for 0, ±2%, ±4 and ±6% ($\Delta V_0/V_0$) uniformly strained CoFeCrGa structures. The solid red and blue lines show the majority and minority DOS. The Fermi energy level is shifted to zero.

Concerning the magnetization of the cubic strained structures, the total magnetic moment remains the same as in the *Y-I* ordered structure (≈ 2.00 $\mu_B$/*f.u.*) due to their analogous DOS plots, as shown in Figure 1(d). Regarding the atomic magnetic moments (AMMs), Cr-AMM increases, Fe-AMM decreases, and Co-AMM remains nearly constant, with respect to *the Y-I ordered structure*, as the strain values increase. Since the Ga-AMM is very small, consistent with the general observation that the *sp*-group elements show a diamagnetic character in HAs, it is not represented in Figure 1(d). The increase in Cr-AMM compensates for the decrease in Fe-AMM and interstitial moments, leaving the total spin moment for the strained structures constant as in the ideal structure, across all uniform strain values. Since the DOS, *P,* and AMM – are all interrelated through the partial (atomic) density of states (PDOS), therefore, the observed variations in AMMs can be attributed to the same physical mechanism that governs the change in the DOS and *P*, *i.e.*, altered exchange interactions and the interatomic hybridizations with strain value [47]. In summary, these results demonstrate that uniform strain did not affect the SGS nature of CoFeCrGa, had a minimal (or no) impact on $M_S$, and is highly likely to happen experimentally.

### 3.2.2 Effect of tetragonal distortion on the structural, electronic, and magnetic properties of CoFeCrGa:

This section discusses the effect of tetragonal distortion on the structural, electronic, and magnetic properties of CoFeCrGa alloy. Figures 3 & 4 present the structural, electronic, and magnetic properties of tetragonally deformed structures with different axial ratios (*c/a* values, with $V_0$ unit-cell volume). As presented in Figure 3(b), all tetragonally deformed structures have positive RFEs; therefore, the formation of a tetragonally distorted structure is less favorable compared to the *Y-I* ordered structure.

For the tetragonally distorted structures with very small distortion values, particularly having *c/a* = 0.90 to *c/a* = 1.20, the RFE is exceedingly small (≤ 0.1 eV/*f.u.*), thus making their experimental occurrence highly probable. Beyond these *c/a* values; RFE increases rapidly with increasing $|\Delta c/a|$ values, resulting in a decreased probability of occurrence for these deformed structures. Another noteworthy observation from the RFEs of the tetragonally distorted structures is that RFE is considerably lower for the tetragonally distorted structures with $c/a \geq 1.0$ compared to those with $c/a \leq 1.0$. This asymmetric change in RFE arises from the interplay of attractive and repulsive interatomic forces within the distorted structures. Thus, it is concluded that the favored condition for the occurrence of tetragonal distortion within CoFeCrGa is elongation along the *c*-axis (with compression of the *ab*-plane) rather than compression along the *c*-axis (with elongation of the *ab*-plane).

Following that, let us discuss the DOS and *P* for the tetragonally distorted structures. For the tetragonally distorted structures, the DOS undergoes significant modifications as a consequence of the reduced crystal symmetry. Microscopically, the variations in the PDOS (and consequently in the total DOS) originate from changes in atomic exchange interactions and interatomic hybridization—similar to the case of uniformly strained structures. However, owing to the change in lattice symmetry at the macroscopic level, these modifications are more pronounced in the tetragonally distorted phases. Consequently, the key features of the SGS DOS *i.e.*, peak and valley characteristics and spin gapless point, are lost; and a smoother DOS is observed for all tetragonally distorted structures. This phenomenon is clearly evident in their DOS plots, as depicted in Figure 4. For the slightly deformed structures ($0.90 \leq c/a \leq 1.10$), there are small changes in DOS, and the DOS plots exhibit a nearly half-metallic nature due to the presence of an energy gap in the minority spin channel around $E_F$. This results in ≈ 90% spin polarization for them, akin to half-metallic compounds.

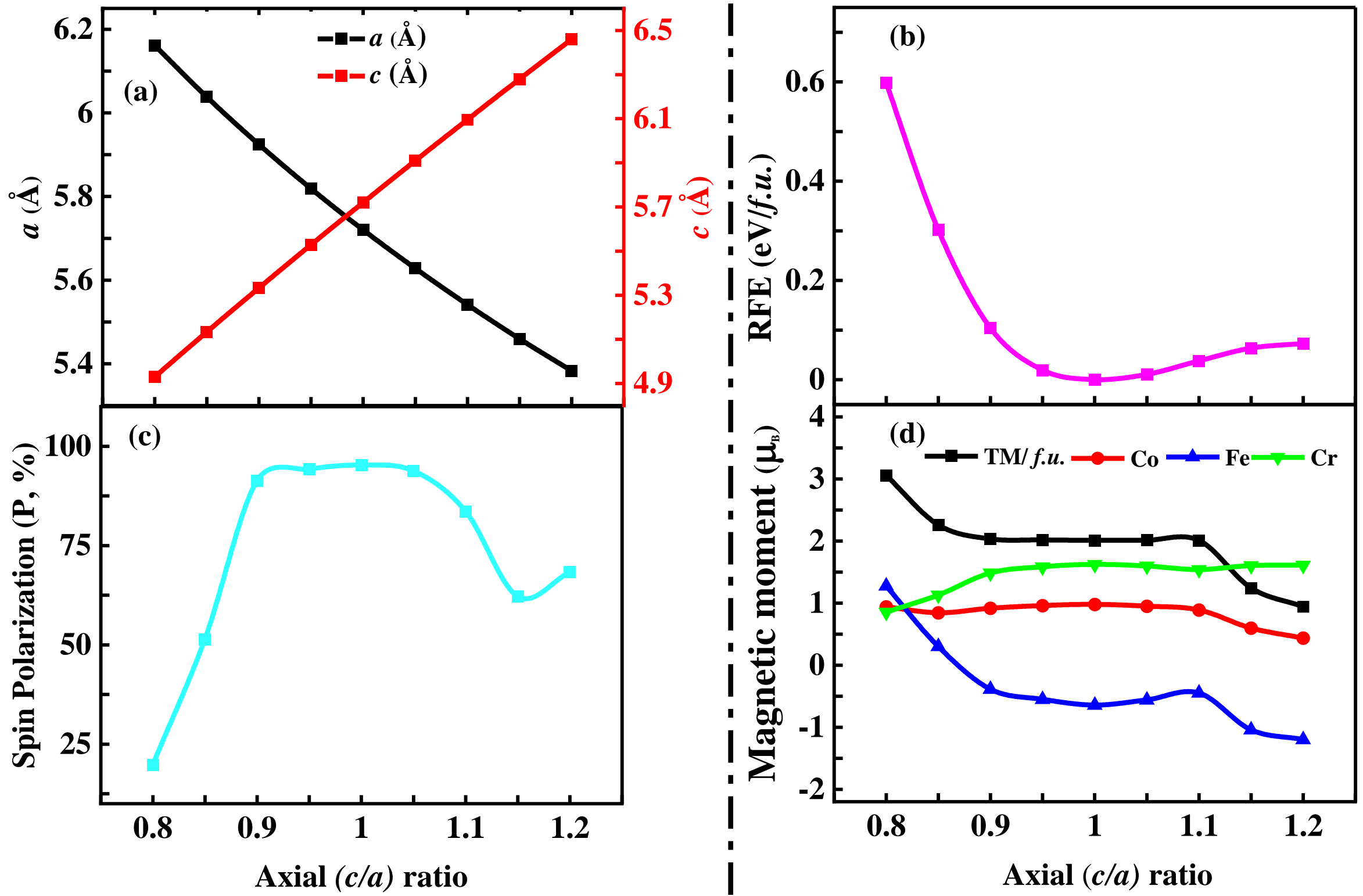

**Figure 3**: (a) Lattice parameters, (b) Relative formation energy (RFE), (c) spin polarization, (d) total and atomic magnetic moments for tetragonally distorted CoFeCrGa alloys as a function of *c/a*.

On the other hand, in the case of the distorted structures with higher distortion ($|\Delta c/a|$) values, DOS becomes smoother and more dispersive, accompanied by a complete closure of the minority energy gap (due to the substantially increased atomic orbitals' overlapping along the contraction directions). Therefore, the deformed structures with higher distortion values (*i.e.*, with $|\Delta c/a| \geq 0.10$) demonstrate a dominating metallic nature, accompanied by a substantial reduction in the $P$. For these structures with higher distortion values, the spin polarization declines almost monotonically with increasing distortion ($\Delta c/a$) values and $P$ reduced to ~25% (at $c/a = 0.8$) and ~75% (for $c/a = 1.2$), as illustrated in Figure 3(c). The key factor governing this trend in $P$ is the modified DOS shape for the tetragonally deformed structures. Moreover, the decline in $P$ with increasing $|\Delta c/a|$ values for tetragonally deformed structures can be ascribed to the weakening of the exchange interactions and interatomic hybridization [11]. In summary, tetragonally deformed structures with smaller distortion values ($|\Delta c/a| \leq 0.10$) display a nearly half-metallic nature with very high spin polarization (≈ 90%), whereas those with the larger distortion values display a dominating metallic character with a significantly reduced spin polarization, depending on their axial ratio.

The total magnetic moment and AMMs follow a similar trend that of spin polarization. For small distortion ($0.90 \leq c/a \leq 1.10$), the total magnetic moment does not change too much, and the distorted structures exhibit nearly the same total moment values as of *Y-I* ordered structure (~2.0 $\mu_B$/*f.u.*). A further large deviation of *c/a* from 1.0 leads to a significant and arbitrary alternation in magnetic moment values, as depicted in Figure 2(d), such as for the distorted structures with *c/a* = 0.8 and 1.2, the total moment shoots to 3.05 $\mu_B$/*f.u.* and 0.95 $\mu_B$/*f.u.* A similar trend is also followed by AMMs. These arbitrary changes in the AMMs and total magnetic moments arise from the alterations in the constituent atoms' PDOS and total DOS of distorted CoFeCrGa structures, respectively.

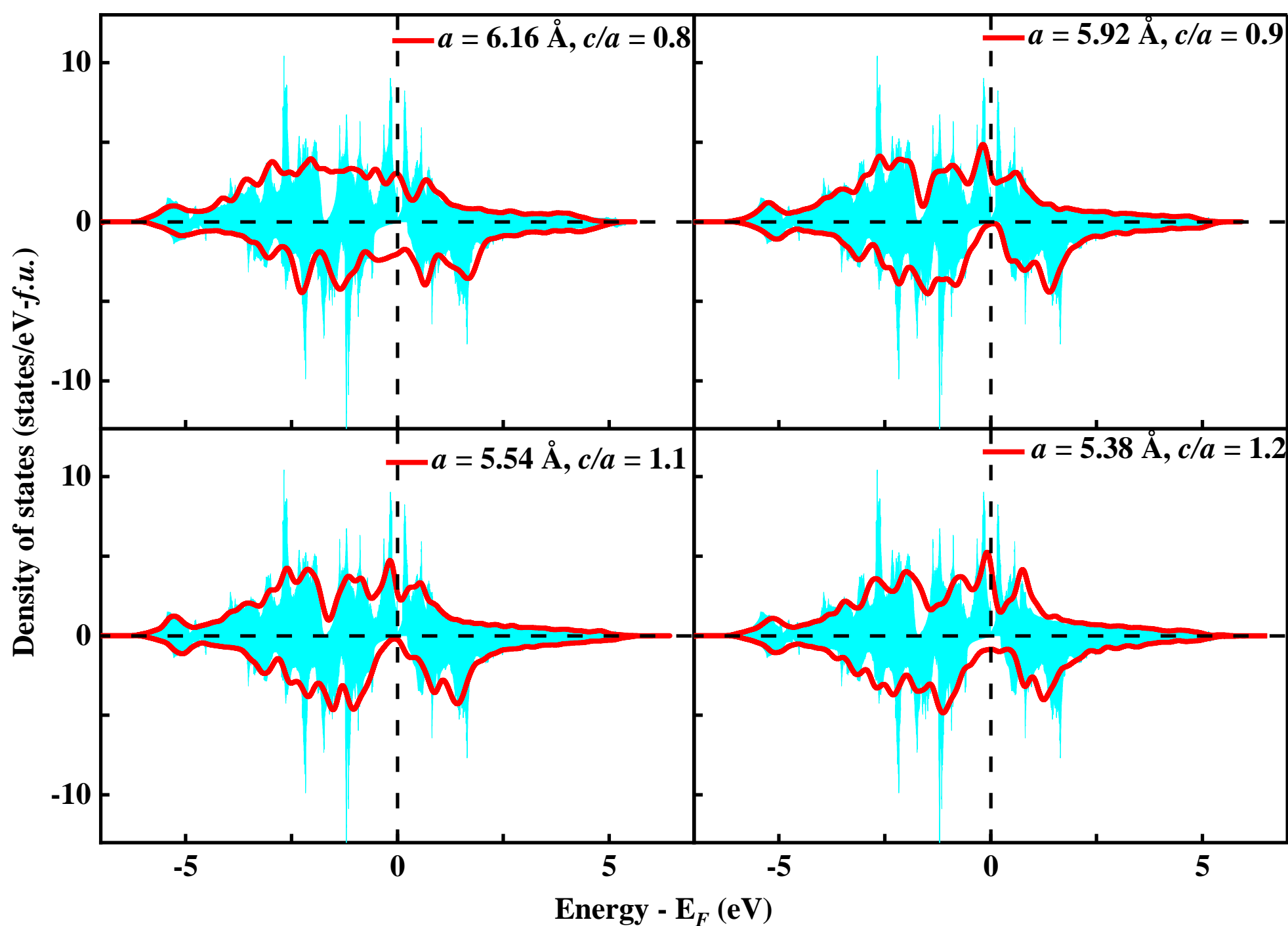

**Figure 4**: The density of states plots for tetragonally distorted CoFeCrGa alloy with axial ratio (*c/a*) of 0.8, 0.9, 1.1 & 1.2. The Fermi energy level is shifted to zero in all plots.

Finally, we discuss the structural stability of the uniformly strained structures and the tetragonally distorted structures. The RFE provides information not only about the ease of occurrence of disorder formation but also about the relative thermodynamic stability of the defective structures w.r.t. the ideal reference structure. For the uniformly strained structures, RFE is a very small positive value, indicating these structures are metastable. On the other hand, the tetragonally distorted CoFeCrGa structures exhibit much larger positive RFE, implying that they are considerably less stable than the Y-I ordered structure. The underlying reason for the reduced stabilities of the tetragonally distorted CoFeCrGa structures lies in their electronic configurations, mainly attributable to the higher valence band energies in the tetragonally deformed structures compared to that of the optimized *Y-I* structure. The valence band energy for an electronic structure is given by $E_{band} = \int_{E=Ev}^{EF} \mathrm{d}E * DOS(E) * E$, as discussed in ref. [15]. From Figure 4, it becomes evident that the tetragonally deformed structures have higher DOS in the proximity of $E_F$, compared to the optimized *Y-I* structure. Hence, in the conjugation with the total number of valence electrons $N_v = \int_{E=E_v}^{E_F} dE\, DOS(E)$, the increased DOS around $E_F$ for the tetragonally distorted structure leads to higher band energies than the Y-I structure, making the tetragonally distorted structures less stable compared to the *Y-I* ordered structure. Notably, this observation is also consistent with the findings of Matsushita *et al.* [14] and Faleev *et al.* [13] for other HAs, which reported that only the cubic HAs with DOS ≥ 4.5 states/eV and 3.0 states/eV at $E_F$ respectively, tend to form stable or metastable tetragonal phase (through a reduction of sharp DOS peaks near $E_F$). This suggests that the tetragonal phase of CoFeCrGa is not excessively unstable and may be accessible experimentally through non-equilibrium deposition techniques.

### 3.3 Lattice distortion-induced magnetic anisotropy

Ferromagnetic materials with high magnetic anisotropy are crucial for various spintronic applications, and by assessing magnetic anisotropy, one can select the material suited for the specific technological applications. For example, the materials with a high magnetic anisotropy on the order of $10^6 – 10^7$ J/m$^3$, coupled with a magnetic hardness parameter ≥ 1.0 and a high Curie temperature, are ideal for permanent magnets and conventional storage devices. Similarly, for perpendicular magneto-resistive random-access memories, magnetic materials with perpendicular magnetic anisotropy of $10^5 – 10^6$ J/m$^3$ coupled with ~100% spin polarization and high Curie temperature are preferable [48–50]. Hence, the knowledge of magnetic anisotropy energies (MAE) of a particular material is highly valuable for designing spintronics devices.

Numerous studies have shown that lattice deformation can induce magnetic anisotropy within the ferromagnetic material, due to the differing crystal field and demagnetization field along the different axes of the distorted unit cell [14,49,51–54]. Consequently, it is expected that the lattice-deformed CoFeCrGa cubic unit cells may also exhibit magnetic anisotropy. In search of possible magnetic anisotropy (MA) for these structures, we focus on two most relevant types of MA: magneto-crystalline anisotropy (MCA), and magnetic shape anisotropy (MSA) aka magnetic dipolar anisotropy or uniaxial anisotropy (as known among the experimentalists). These types of anisotropy are anticipated to be most probably present in the lattice-deformed cubic FM structures, as suggested by the literature on other FM materials. Consequently, the total MA will be the summation of MCA and MSA. The discussion will first address MCA, followed by MSA and total MA.

The MCA for the uniformly strained and tetragonally deformed CoFeCrGa structures have been investigated by utilizing the magnetic force theorem as implemented in QUATUM ESPRESSO package [55–57]. The calculated MCA is presented in two forms – the total MCA energy per *f.u.* (MCA) and the total MCA energy per unit-cell volume (which is also known as MCA constant ($K_{MCA}$)). These quantities are calculated as follows –

$$\begin{gathered} MCA = E_{band}[100] - E_{band}[001] \\ K_{MCA} = MCA/V \end{gathered} \tag{1}$$

Here, $E_{band}$ refers to the fully relativistic band energies (including SOC), obtained from non-self-consistent calculations but initialized using a well-converged spin density obtained using a collinear self-consistent calculation for computational simplicity, and V is the unit-cell volume for the lattice-deformed structure. In the present study, the total MCA energy is calculated as the difference between the fully relativistic band energies along the [100] and [001] directions, as also reflected from equation (1). These directions are selected for calculations because the most pronounced effects of crystal field (CF), spin–orbit coupling (SOC), and exchange interactions (EI)—the key factors determining MCA—are expected to be extremal along these directions in lattice-distorted CoFeCrGa structures. Thus, one of these directions corresponds to the easy axis and the other to the hard axis of magnetization. According to equation (1), a positive MCA value indicates that the out-of-plane magnetization configuration is energetically favorable (or the magnetization easy axis is along [001]) arising from MCA. Whereas a negative value of MCA indicates that the in-plane magnetization configuration is energetically favorable (or magnetization easy axis is along [100]). Since MCA is highly sensitive to the *k*-points density employed for calculations due to its relatively small magnitude, the accuracy of the calculated MCA is validated by performing calculations with a finer *k*-points mesh for several tetragonally deformed structures, which are equivalent to $(17)^3$ *k*-points mesh in the *Y-I* ordered structures [$(5.72\ \text{Å})^3$ cell-size]. This refinement results in only a small change ($\leq 0.01$ meV/*f.u.*) in the total MCA. Therefore, MCA for all tetragonally deformed structures reported here is calculated with a *k*-point mesh equivalent to $(15)^3$ *k*-points mesh in the *Y-I* ordered structure, while maintaining the same *k*-points density. A much higher *k*-points grid has not been tested due to computational resource limitations.

For the Y-I ordered and isotopically distorted CoFeCrGa structures, the calculated MCA is nearly zero (a few μeV/atom, within the numerical accuracy). This is because magnetic anisotropy originates from the combined effects of electrostatic crystal-field interactions, spin-orbit coupling, and exchange interactions, as discussed earlier. In the cubic CoFeCrGa structures, the orbital field is almost quenched by the symmetrical crystal field, thereby suppressing MCA. Consequently, the *Y-I* ordered and uniformly strained CoFeCrGa structures exhibit negligible MCA. On the other hand, under the tetragonal distortion, the cubic lattice symmetry of CoFeCrGa is broken, leading to the asymmetric crystal field, which only partially quenches the orbital field. As a result, the tetragonally deformed structures show finite and significant MCA [58]. The computed MCA values are presented in Figure 5(a) in units of meV/*f.u.* and J/m$^3$. The out-of-plane compressed structures ($c/a < 1.0$) yield negative MCA, ranging from $-0.28\times10^6$ J/m$^3$ ($c/a = 0.95$) to $-0.17\times10^6$ J/m$^3$ ($c/a = 0.80$), with an extremum value of $-2.07\times10^6$ J/m$^3$ for the distorted structure with axial ratio ($c/a$) of 0.90. This indicates that the compressed structures along the *c*-axis will exhibit the in-plane MCA. In contrast, MCA switches to a positive value for tensile strain ($c/a > 1.0$), and monotonically increases from $1.01\times10^5$ J/m$^3$ ($c/a = 1.05$) to $1.64\times10^6$ J/m$^3$ ($c/a = 1.2$) with the increasing $c/a$ value, indicating that the elongated structures along the *c*-axis will have out-of-plane MCA. As a result, MCA displays a bipolar characteristic with respect to the distortion value ($\Delta c/a$). Following the method discussed in ref. [57], the atomic- and orbital-resolved contributions to the total MCA are evaluated for all lattice-deformed structures using equation (2), to elucidate the microscopic origin of MCA.

$$MCA_\alpha = \int^{E_F} (E - E_F) n_\alpha^{[100]}(E) dE - \int^{E_F} (E - E_F) n_\alpha^{[001]}(E) dE; \quad \mathrm{MCA}_\alpha = \sum_\beta \mathrm{MCA}_{\alpha\beta} \tag{2}$$

Here, α and β stand for the atoms and orbitals, respectively. Since the calculated MCA for the *Y-I* ordered and uniformly strained CoFeCrGa structures are almost zero, their corresponding atomic and orbital resolved MCA are also very small, on the order of ~μeV/atom, and are therefore not discussed further. In contrast, for the tetragonally distorted structures ($c/a \neq 1$), the atomic and orbital contributions to MCA are of the order of ~meV/atom, and therefore, are presented in Figures 5(b) and 6. As shown in Figure 5(b), among all tetragonally distorted structures, those with axial ratios (*c/a*) of 0.90, 1.15, and 1.20 have unidirectional atomic-resolved MCA, leading to significantly large total MCA, including the extremum at *c/a* = 0.90 and 1.20, for negative and positive MCA, respectively. For the rest of the distorted structures, the atomic-resolved MCA contributions are bi-directional, yielding some intermediate values for total MCA.

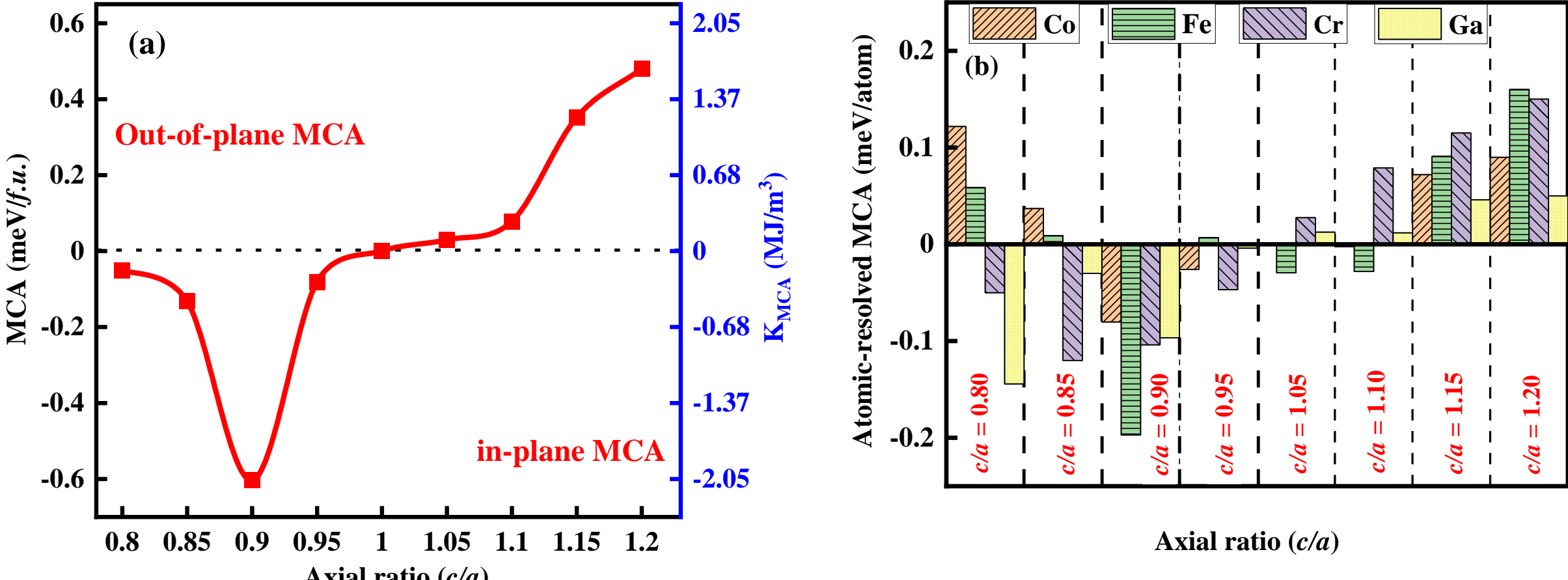

**Figure 5**: (a) Total MCA for tetragonally deformed CoFeCrGa, as a function of their axial ratio (*c/a* value), (b) atomic contributions to MCA (atomic-resolved MCA) in the case of the tetragonally distorted structure with *c/a* = 0.8 to *c/a* = 1.2 (left to right panels).

The orbital-resolved MCA highlights that the outer-shell orbitals of constituent atoms (*i.e.*, *d*-orbitals of transition metals (TMs) and the *p*-orbitals of the Ga atoms) contribute most significantly to the total MCA. In contrast, the inner-shell orbitals of the respective atoms (*i.e.*, *s*- and *p*-orbitals for the *TMs*, and *s*-orbitals for Ga) make only a negligible contribution. This is reasonable, since the effects of (CF+SOC+EI) are strongest for TM-*d* orbitals and Ga-*p* orbitals, whereas the fully filled inner-shell orbitals experience minimal SOC effects. However, the inner orbitals still contribute, though their contribution is small, due to their hybridization with the outer shell orbitals. In addition, minor contributions also arise from the interstitial region due to orbital interactions in that space. Because both the inner-shell and interstitial contributions are much smaller than those from the outer-shell orbitals, only the TM-*d* and Ga-*p* orbitals' contributions are shown in Figure 6, as they dominate the overall MCA. For example, for the distorted structure with $c/a$ = 0.90, the major contribution to MCA arises from the Co-$d_{z^2}$, Co-($d_{x^2-y^2}$), Fe-$d_{z^2}$, Cr-$d_{zx}$, Cr-$d_{zy}$, and Ga-$p_y$ orbitals; whereas for the most elongated structure along the $c$-axis (*i.e.*, structure with $c/a$ = 1.20), the dominant contributions originate from Cr-($d_{x^2-y^2}$), *Fe-*$d_{xy}$, *Fe-*($d_{x^2-y^2}$), and *Ga-*$p_y$ orbitals. The orbital-resolved MCA for other distorted structures exhibits similar variations, as illustrated in Figure 6. These variations in orbital-resolved MCA are closely linked to the redistribution of their electronic states in the vicinity of $E_F$ [59,60]. Furthermore, with the help of the orbitals resolved MCA, the comparable atomic contributions to MCA for TM-atoms and Ga-atoms, for the distorted structures with the axial ratios between 0.8 and 0.9, can be explained. Their comparable contributions may initially appear anomalous given their different (CF+SOC+EI) profiles. However, the detailed orbital-resolved MCA analyses (Figure 6) revealed that the absolute contribution from TMs-orbitals to the total MCA is much larger than that of Ga-orbitals. However, these TMs-orbitals' contributions tend to offset one another as a result of their opposing character (negative for some orbitals and positive for others). Consequently, the net MCA contribution from the TMs appears comparable to that of Ga.

**Orbital-resolved magneto-crystalline anisotropy**

Axial ratio ($c/a$)

**Co-3*d* orbitals' MCA (meV)**

| Axial ratio | $d_{z^2}$ | $d_{zx}$ | $d_{zy}$ | $d_{x^2-y^2}$ | $d_{xy}$ |
|---|---|---|---|---|---|
| $c/a$ = 1.20 | 0.012 | -0.010 | 0.021 | 0.030 | 0.033 |
| $c/a$ = 1.15 | 0.002 | -0.009 | 0.022 | 0.042 | 0.009 |
| $c/a$ = 1.10 | -0.101 | -0.009 | -0.012 | 0.085 | 0.030 |
| $c/a$ = 1.05 | -0.105 | -0.010 | -0.016 | 0.101 | 0.024 |
| $c/a$ = 0.95 | -0.101 | 0.006 | -0.014 | 0.087 | 0.004 |
| $c/a$ = 0.90 | -0.108 | 0.034 | -0.017 | 0.070 | 0.003 |
| $c/a$ = 0.85 | -0.128 | 0.058 | 0.030 | 0.077 | 0.026 |
| $c/a$ = 0.80 | -0.149 | 0.065 | 0.050 | 0.106 | 0.112 |

**Cr-3*d* orbitals' MCA (meV)**

| Axial ratio | $d_{z^2}$ | $d_{zx}$ | $d_{zy}$ | $d_{x^2-y^2}$ | $d_{xy}$ |
|---|---|---|---|---|---|
| $c/a$ = 1.20 | -0.028 | 0.011 | -0.005 | 0.146 | 0.020 |
| $c/a$ = 1.15 | -0.034 | 0.010 | -0.007 | 0.122 | 0.018 |
| $c/a$ = 1.10 | -0.018 | 0.008 | -0.001 | 0.089 | -0.001 |
| $c/a$ = 1.05 | -0.042 | 0.006 | -0.001 | 0.066 | -0.003 |
| $c/a$ = 0.95 | -0.065 | -0.009 | -0.004 | 0.032 | -0.001 |
| $c/a$ = 0.90 | -0.101 | -0.026 | -0.012 | 0.033 | 0.017 |
| $c/a$ = 0.85 | -0.103 | -0.035 | -0.004 | 0.022 | 0.004 |
| $c/a$ = 0.80 | -0.093 | -0.029 | 0.006 | 0.033 | 0.040 |

**Fe-3*d* orbitals' MCA (meV)**

| Axial ratio | $d_{z^2}$ | $d_{zx}$ | $d_{zy}$ | $d_{x^2-y^2}$ | $d_{xy}$ |
|---|---|---|---|---|---|
| $c/a$ = 1.20 | 0.031 | -0.003 | -0.013 | 0.048 | 0.090 |
| $c/a$ = 1.15 | 0.015 | 0.009 | -0.008 | 0.038 | 0.019 |
| $c/a$ = 1.10 | -0.061 | 0.008 | -0.030 | 0.028 | 0.020 |
| $c/a$ = 1.05 | -0.077 | 0.009 | -0.038 | 0.045 | 0.027 |
| $c/a$ = 0.95 | -0.034 | -0.006 | -0.035 | 0.048 | 0.039 |
| $c/a$ = 0.90 | -0.034 | -0.080 | -0.079 | 0.016 | 0.035 |
| $c/a$ = 0.85 | -0.054 | -0.001 | -0.011 | 0.094 | 0.045 |
| $c/a$ = 0.80 | -0.100 | 0.054 | 0.083 | 0.152 | 0.049 |

**Ga-4*p* orbitals' MCA (meV)**

| Axial ratio | $p_z$ | $p_x$ | $p_y$ |
|---|---|---|---|
| $c/a$ = 1.20 | -0.001 | 0.006 | 0.041 |
| $c/a$ = 1.15 | -0.005 | 0.007 | 0.044 |
| $c/a$ = 1.10 | -0.012 | 0.002 | 0.023 |
| $c/a$ = 1.05 | -0.012 | 0.002 | 0.022 |
| $c/a$ = 0.95 | 0.010 | -0.002 | -0.011 |
| $c/a$ = 0.90 | 0.038 | -0.032 | -0.108 |
| $c/a$ = 0.85 | 0.007 | -0.011 | -0.042 |
| $c/a$ = 0.80 | -0.060 | -0.024 | -0.090 |

Color scale: 0.16, 0.12, 0.08, 0.04, 0, -0.04, -0.08, -0.12, -0.16

**Figure 6**: The orbital resolved MCA for different tetragonally deformed structures. The $y$-axis of maps represents the $c/a$ value for the tetragonally deformed structures, whereas the $x$-axis corresponds to the different $d$- and $p$-orbitals. The magneto-crystalline anisotropy values are shown on *the z-axis, with a common color weight given on the right side of the* plots.

A final comment on the MCA of the tetragonally distorted structures concerns the non-monotonic behavior of MCA under negative tetragonal distortion ($\Delta c/a < 0$). The extremum hump at $c/a = 0.90$ may seem unexpected, as one might expect monotonic change for MCA with increasing distortion degree, as observed under positive tetragonal distortion ($\Delta c/a > 0$). Although most materials exhibit a monotonic change in MCA with distortion value [61,62], non-monotonic trends have also been reported for some magnetic materials [63–65]. Such behavior can arise from several reasons, including the complex interaction between structural and magnetic parameters, the intrinsic characteristic of magnetic material, or the presence of additional magnetization directions with extremum band energies beyond [100] and [001]—the only directions considered for MCA calculations in this work. Thereby, to fully capture the non-monotonic response of MCA under negative tetragonal distortion, a more rigorous analysis incorporating additional magnetization directions might be helpful, since [100] and [001] directions may not correspond to the true axes of extremum band energies in the tetragonally deformed CoFeCrGa structures.

After that, we discuss the magnetic shape anisotropy of lattice-distorted CoFeCrGa structures, which arises from the non-spherical geometry of the magnetic material. The MSA for bulk lattice-distorted CoFeCrGa structures is calculated as the difference of magnetic dipolar energy ($E_{dip}$) between the [100] and [001] directions –

$$MSA = E_{dip}[100] - E_{dip}[001], \qquad K_{MSA} = MSA/V \tag{2}$$

Analogous to MCA, the calculated MSA is reported both as the total MSA energy per *f.u.* (MSA) and as the total MSA energy per unit-cell volume, which is also known as MSA constant ($K_{MSA}$). Here, $E_{dip}$ is calculated using the direct summation method, following refs. [66] and [67,68], and is given by the following equation:

$$\boldsymbol{E_{dip}} = -\frac{1}{2}\sum_{i} \vec{m}_i \cdot \vec{B}_i, \qquad \vec{B}_i = \frac{\mu_0}{4\pi}\sum_{j} \frac{3(\vec{m}_j \cdot \hat{r}_{ij})\hat{r}_{ij} - \vec{m}_j}{|\vec{r}_{ij}|^3} \tag{3}$$

Here, $i$ runs over the magnetic atoms in unit-cell, while $j$ runs over all atoms within a large cut-off sphere with a radius of 200 Å (see section-II of the Supplemental material for the choice of cut-off sphere radius). Similar to the MCA calculations, two directions for magnetization – [100] and [001] – are considered for the MSA calculation, as the maximum variation in the dipolar energy of the distorted structure is anticipated between these axes. As a result, the positive MSA corresponds to an out-of-plane easy axis, and the negative MSA corresponds to an in-plane easy axis for magnetization - all arising from MSA. Finally, the MSA of the lattice-distorted structures is compared with the corresponding MCA, to find out the overall magnetic anisotropy and preferred magnetization orientation.

For the cubic structures (*Y-I* ordered and uniformly strained structures), the calculated MSA is zero due to similar dipolar energies along [100] and [001] directions, a consequence of their cubic (nearly spherical) geometry. Consequently, both *Y-I* ordered and uniformly strained structures exhibit magnetic isotropic nature, as both MCA (as discussed earlier) and MSA are *zero* for them. On the other hand, for the tetragonally deformed structures, significant anisotropic geometry leads to different dipolar energies along [100] and [001], resulting in non-zero MSA. Thereby, in Figure 7, MSA is shown only for the tetragonally deformed structures and compared with corresponding MCA values. However, the calculated MSA for these structures is on the order of $\sim 10^4$ J/m$^3$, and is much smaller than the MCA for the corresponding structures. These low MSA values can be attributed to the small distortion, as well as to the co-existence of positive and negative atomic magnetic moments within the distorted structures (as seen in Figure 3(d)) [60].

As a result, total MA will be determined by MCA, and MSA have a negligible impact. Therefore, in the overall picture, the in-plane compressed ($c/a < 1.0$) structures will exhibit the in-plane magnetic anisotropy, with the preferred magnetization direction along [100]. On the other hand, the out-of-plane elongated ($c/a > 1.0$) structures will exhibit the out-of-plane-magnetic anisotropy (OMA), or perpendicular magnetic anisotropy (PMA), with the preferred magnetization direction along [001]. Notably, PMA is highly desirable for spintronic devices, as it enables higher data storage density with lower switching current and improved thermal stability, compared to those materials with in-plane anisotropy.

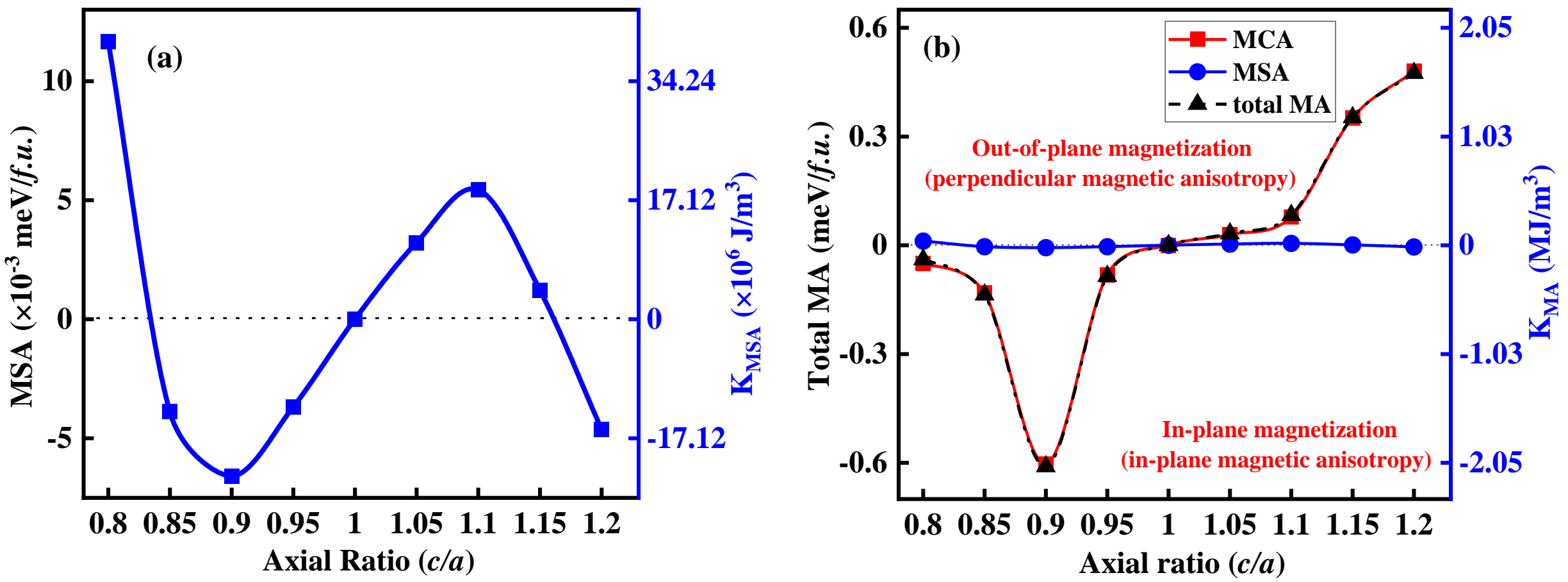

**Figure 7**: (a) The MSA for tetragonally deformed structures calculated using equation 3, (b) the total MA for the tetragonally deformed structures, considering the contribution from MCA and MSA.

An important and final observation regarding the MA in CoFeCrGa is that the total MA in tetragonally distorted structures exhibits a well-defined bipolar nature. This bipolar nature of total MA suggests that by inducing the tetragonal distortion *via* the controlled preparation method or controlled growth coupled with a suitable adjacent layer, the desirable magnetic orientation for CoFeCrGa can be easily achieved. For example, the non-equilibrium deposition techniques such as sputtering, pulsed laser deposition, or molecular beam epitaxy can easily promote the growth of metastable states for CoFeCrGa with the tetragonal deformed lattice geometry, which, in turn, result in significant MA corresponding to the distortion value. Similarly, depositing the CoFeCrGa along with GaAs (a popular substrate material and as well as the widely used spacer layer for spin valves) and Si (a common substrate material for spintronic devices) may induce uniaxial tensile strain along the out-of-plane direction of CoFeCrGa (as $a_{GaAs} \leq a_{CoFeCrGa}$ and $a_{Si} \leq a_{CoFeCrGa}$), and thereby promoting the PMA within CoFeCrGa. On the other hand, the growth of bulk-CoFeCrGa on MgO or $MgAl_2O_4$ (used as both substrate and the insulating layer materials for magnetic tunnel junctions) is expected to induce IMA within CoFeCrGa, due to the induce compressive strain in out-of-plane direction. Nonetheless, these predictions are solely based on the lattice mismatch of CoFeCrGa with adjacent layers, assuming that the CoFeCrGa unit-cell volume is not changed after the lattice deformations; as observed in similar material $Co_2MnAl$, where tetragonal deformations with the optimized cubic cell volume are most stable [7]. Furthermore, the calculated MCA, and hence MA of the lattice deformed CoFeCrGa structures, is higher than most of the other Co-based full Heusler alloys, which have MA values range from $-0.09\times10^6$ J/m$^3$ to $-0.31\times10^6$ J/m$^3$ [69]. Such high magnetic anisotropy ensures the high thermal stability and high storage density of the spintronic devices fabricated using CoFeCrGa.

Thus, in summary, the very high values and well-established bipolar nature of MA with distortion values indicate that the desired magnetic orientation of CoFeCrGa can be readily tailored for specific applications.

### 3.4 Lattice distortion-induced Hall conductivies

Transverse Hall effects are of particular interest for the development of next-generation spintronic devices, owing to the inherently dissipationless nature of the transverse currents. These effects enable a wide range of applications, including high-sensitivity magnetic field sensors, read-out heads, spin–orbit torque-based magneto-resistive random-access memories, and terahertz emitters. Many Co-based full Heusler alloys, such as $Co_2MnAl$, $Co_2MnGa$, and $Cu_2CoSn$, have been reported to exhibit very large anomalous Hall conductivity (AHC) and spin Hall conductivity (SHC), due to their large Berry curvature distribution. In these materials, broken time-reversal symmetry, spin–orbit coupling, and crystallographic symmetries are key factors that generate high Berry curvature (BC) through the formation of strong hotspots, leading to significantly enhanced AHC and SHC [70]. Therefore, investigating the Hall conductivities of CoFeCrGa is anticipated to yield valuable insights, as it may reveal similar mechanisms governing its transverse transport properties. Furthermore, lattice distortions have been reported to modify both AHC and SHC for many magnetic materials through the changes in the Berry curvature profile [71,72]. Consequently, in the present study, the intrinsic AHC and SHC of lattice-distorted CoFeCrGa structures have been systematically examined.

For calculating the AHC and SHC for the lattice distorted CoFeCrGa structures, the tight-binding Hamiltonian is constructed using the maximally localized Wannier functions (MLWF) formed by projecting Bloch wave functions into Co-*d*, Fe-*d*, Cr-*d*, and Ga-*p* orbitals. For all lattice-distorted structures, the electronic band structures obtained from Wannier interpolation are found to reproduce the DFT bands in the energy range of -5 to +5 eV around $E_F$, indicating that the obtained MLWF are suitable for calculating the transport properties. For reference, Figure 8(a) displays the DFT and Wannier-interpolated bands of the Y-I ordered structure (optimized lattice parameter, $a$ = 5.72 Å), which demonstrates excellent agreement between them. A dense 101×101×101 Berry mesh is employed to ensure convergence of the BC, for the evaluation of AHC and SHC. Finally, the intrinsic AHC and SHC are calculated using the Kubo formula. In the Kubo formalism, the AHC calculation is analogous to integrating the charge BC. The SHC is computed in the same manner, simply by substituting the electron velocity operator with the spin current velocity operator [73,74].

The calculated AHC and SHC for the uniformly strained and tetragonally distorted structures are presented in Figures 8(b)-(c). For the Y-I ordered structure, both AHC and SHC are very small, –29.78 S/cm and –8.67 ($\hbar$/e)·S/cm, respectively. The uniformly strained structures exhibit slightly different AHC and SHC compared to those for Y-I ordered structures; however, these modifications are minor, and both HCs remain relatively small across all uniformly strained structures (–50 S/cm ≤ AHC ≤ 80 S/cm) and (–10 ($\hbar$/e)·S/cm ≤ SHC ≤ 90 ($\hbar$/e)·S/cm), indicating that uniform strain does not induce substantial changes in Berry curvature response. In contrast, the tetragonally distorted structures exhibit a pronounced enhancement in AHC, with values ranging from –215 to 250 S/cm. The SHC, however, remains low under tetragonal distortion as well (–10 to 100 ($\hbar$/e)·S/cm), with values comparable to those of the uniformly strained structures. These modifications in AHC and SHC arise from SOC-induced modifications in charge and spin BC, respectively. Overall, these findings demonstrate that the lattice distortions strongly affect AHC and SHC of CoFeCrGa, suggesting that strain engineering effectively tunes the transverse transport properties of CoFeCrGa and renders CoFeCrGa promising for spintronic device applications.

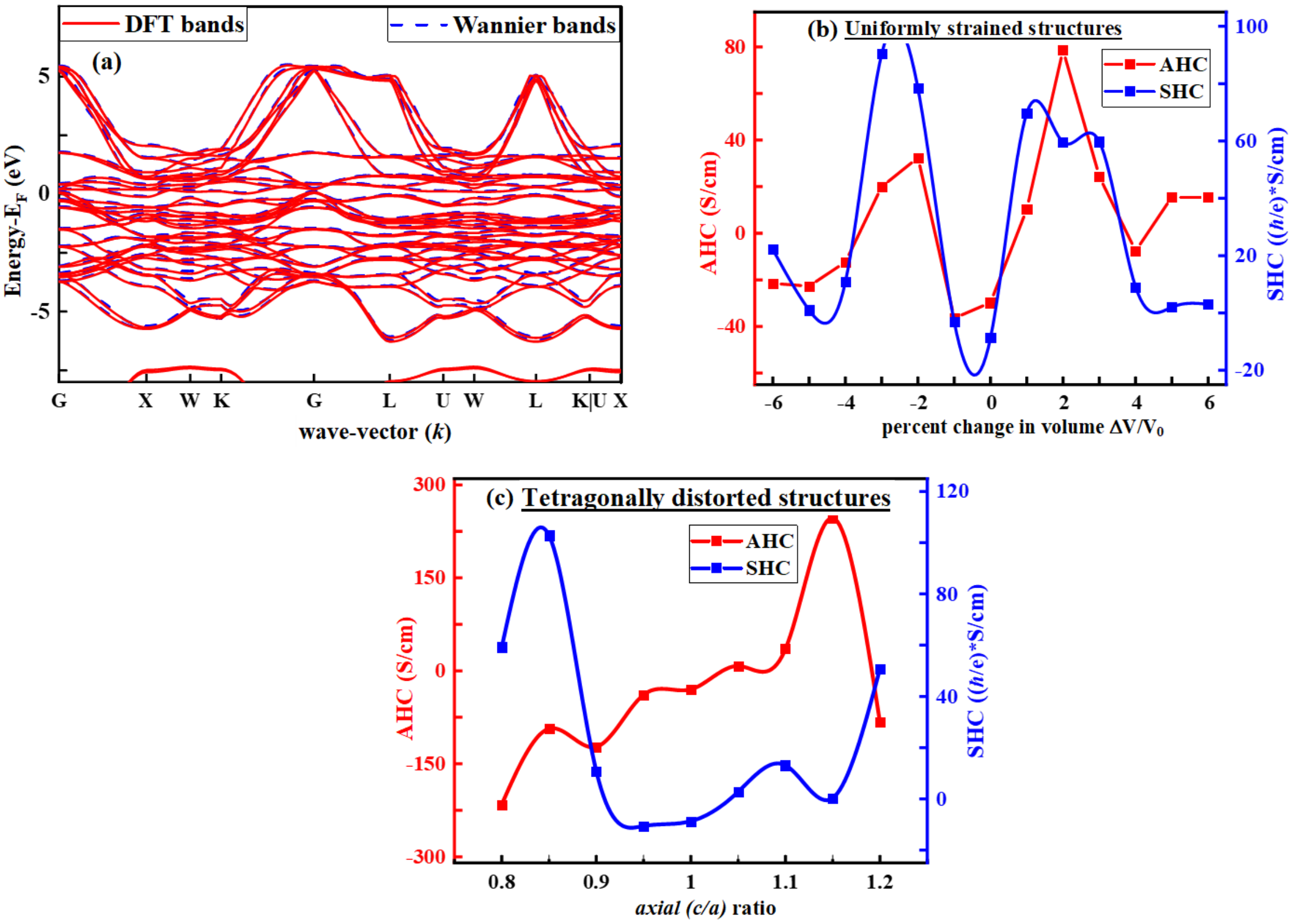

**Figure 8:** (a) Electronic band structure along the high-symmetry points calculated from DFT (red) and the Wannier interpolated band (dashed blue) for the Y-I ordered structure. (b) AHC and SHC for uniformly strained structures, and (c) AHC and SHC for uniformly and tetragonally distorted structures.

## 4. Conclusion

Given that lattice distortions in the constituent materials are readily observable when incorporated into heterostructures or devices, and that the spin gapless semiconducting nature is highly sensitive to external factors, the present study has examined the various physical properties of the uniformly strained and tetragonally distorted CoFeCrGa alloy, with the help of state-of-the-art first-principles-based density functional theory calculations. In particular, the uniform strain in the range -6% $\leq \Delta V/V_0 \leq$ 6% and tetragonal deformation in the range 0.80 $\leq c/a \leq$ 1.20 (at constant $V_0$ unit-cell volume) are considered. This corresponds to a sufficiently wide range of lattice parameters for CoFeCrGa: 5.60 Å $\leq a \leq$ 5.83 Å for the uniformly strained structures and 5.38 Å $\leq a \leq$ 6.16 Å, 4.92 Å $\leq c \leq$ 6.45 Å for the tetragonally distorted structures. The relative formation energy calculations revealed that, for the wide range of distortions, especially the uniformly strained structures correspond to -5% $\leq \Delta V/V_0 \leq$ 6% (lattice parameter: 5.62 Å $\leq a \leq$ 5.83 Å), and tetragonally deformed structures correspond to 0.9 $\leq c/a \leq$ 1.2 (lattice parameters: 5.38 Å $\leq a \leq$ 5.92 Å, 5.33 Å $\leq c \leq$ 6.45 Å) are most likely to be formed experimentally, owing to their very small relative formation energies ($\lesssim$ 0.1 eV/*f.u*). Other distorted structures exhibit larger relative formation energies (RFE), and consequently, their likelihood of occurrence decreases with increasing RFE. Concerning the electronic structure, the SGS nature of CoFeCrGa is robust under uniform strain, with the strained structures exhibiting the integer magnetic moment of 2.0 $\mu_B$/*f.u.*, identical to that of the *Y-I* ordered phase. In contrast, the SGS nature of CoFeCrGa is found to be volatile under the tetragonal (uniaxial) distortions, and even a small distortion leads to the destruction of the SGS nature. The primary reason for the loss of the SGS nature of CoFeCrGa under tetragonal distortion is the lowering of lattice symmetry from cubic to tetragonal, which induces significant alterations in the density of states profile and thereby results in a different electronic nature from SGS. Explicitly, tetragonally distorted structures with small distortions (0.90 $\leq c/a \leq$ 1.10) exhibit a nearly half-metallic nature, with very high spin polarization (> 90%) and nearly constant magnetization (~2.0 $\mu_B$/*f.u.*). Larger distortion ($|\Delta c/a| \geq 0.10$) leads to a dominating metallic nature, accompanied by arbitrary changes in spin-polarization and magnetization from the ideal structure. Furthermore, relative formation energy calculations indicate that elongation along the c-axis (with compression of the ab-plane) is energetically more favorable than compression along the c-axis (with elongation of the ab-plane), suggesting that the former is the preferred mode of tetragonal distortion in CoFeCrGa.

Regarding the magnetic anisotropy of lattice-distorted CoFeCrGa structures, two primary contributors are considered: magneto-crystalline anisotropy (MCA) and magnetic shape anisotropy (MSA). The MCA is calculated through the band energy differences *via* the magnetic force theorem, while MSA is obtained from dipole-dipole interaction energy differences *via* the direct sum method. The *Y-I* ordered and uniformly strained structures show negligible MCA due to the quenching of the orbital field in the cubic crystal field of CoFeCrGa. On the other hand, the tetragonally deformed CoFeCrGa exhibits strong MCA due to their low-symmetric crystal field, with a magnitude on the order of ~$10^5$ – $10^6$ J/m$^3$. Specifically, the calculated MCA is negative for all structures with compression along the out-of-plane direction ($c/a$ < 1.0), reaching a minimum of -2.07×$10^6$ J/m$^3$ for $c/a$ = 0.9; whereas for all structures elongated along the out-of-plane direction ($c/a$ > 1.0), MCA is positive with maxima of 1.64×$10^6$ J/m$^3$ for $c/a$ = 1.2. Regarding MSA, all cubic structures (Y-I ordered and uniformly strained structures) show negligible MSA due to their isotropic geometry. However, in the case of tetragonally distorted structures, the MSA is finite due to their anisotropic geometries, but still, the MSA is small (of the order of ~$10^4$ J/m$^3$) compared to the corresponding MCA. Then, considering both MCA and MSA for the lattice-distorted structures, the cubic CoFeCrGa structures remain magnetically isotropic, like the Y-I ordered structure. In contrast, the tetragonally deformed

CoFeCrGa structures show strong magnetic anisotropic behavior, with the total MA equaling the MCA (~$10^5 – 10^6$ J/m$^3$, with bipolar nature). Therefore, it follows that tetragonal distortion is an essential condition for achieving magnetic anisotropy within bulk-CoFeCrGa, and MCA will dictate their overall magnetic orientation in such tetragonally distorted structures. Thus, by facilitating the well-defined bipolar nature and high values of MCA (and consequently MA); the in-plane or out-of-plane magnetization can be easily obtained within CoFeCrGa, *via* inducing the tetragonal distortion.

With respect to the transverse transport properties, the intrinsic anomalous Hall conductivity (AHC) and spin Hall conductivity (SHC) of CoFeCrGa have been examined. The cubic structures (Y-I ordered and uniformly strained) exhibit relatively small Hall conductivities, with magnitudes of AHC ≤ 90 S/cm and SHC ≤ 90 ($\hbar$/e)·S/cm. In contrast, the tetragonally distorted structures display significantly enhanced AHC ranging from –215 to 250 S/cm. Meanwhile, SHC remains small even under tetragonal distortion, with values ranging from –10 to 100 ($\hbar$/e)·S/cm.

In conclusion, under a wide range of uniform strains, the uniformly strained CoFeCrGa structures retain physical properties similar to those of the Y-I ordered structure, including the preservation of the SGS character, showing that the advantages of the SGS state remain robust under uniform strain. In contrast, under tetragonal distortion, the distorted structures exhibit nearly half-metallic behavior with very large MA and Hall conductivities. All these physical outcomes—whatever under uniform strain or tetragonal distortion – are highly desirable for FM for spintronic applications in different aspects. Thus, CoFeCrGa is a promising candidate for next-generation spintronic devices owing to its tunable and advantageous physical properties. We anticipate future experimental studies to confirm these findings.

## 5. Acknowledgements

The authors thank the IIT Delhi HPC facilities – PADUM and TEJAS (at Physics department) for computational resources. A.K. acknowledges Council of Scientific and Industrial Research (Grant No. 09/086(1356)/2019-EMR-I) India, for the senior research fellowship.

# Supplemental Material

## I. CoFeCrGa - total density of states with GGA-XC functional

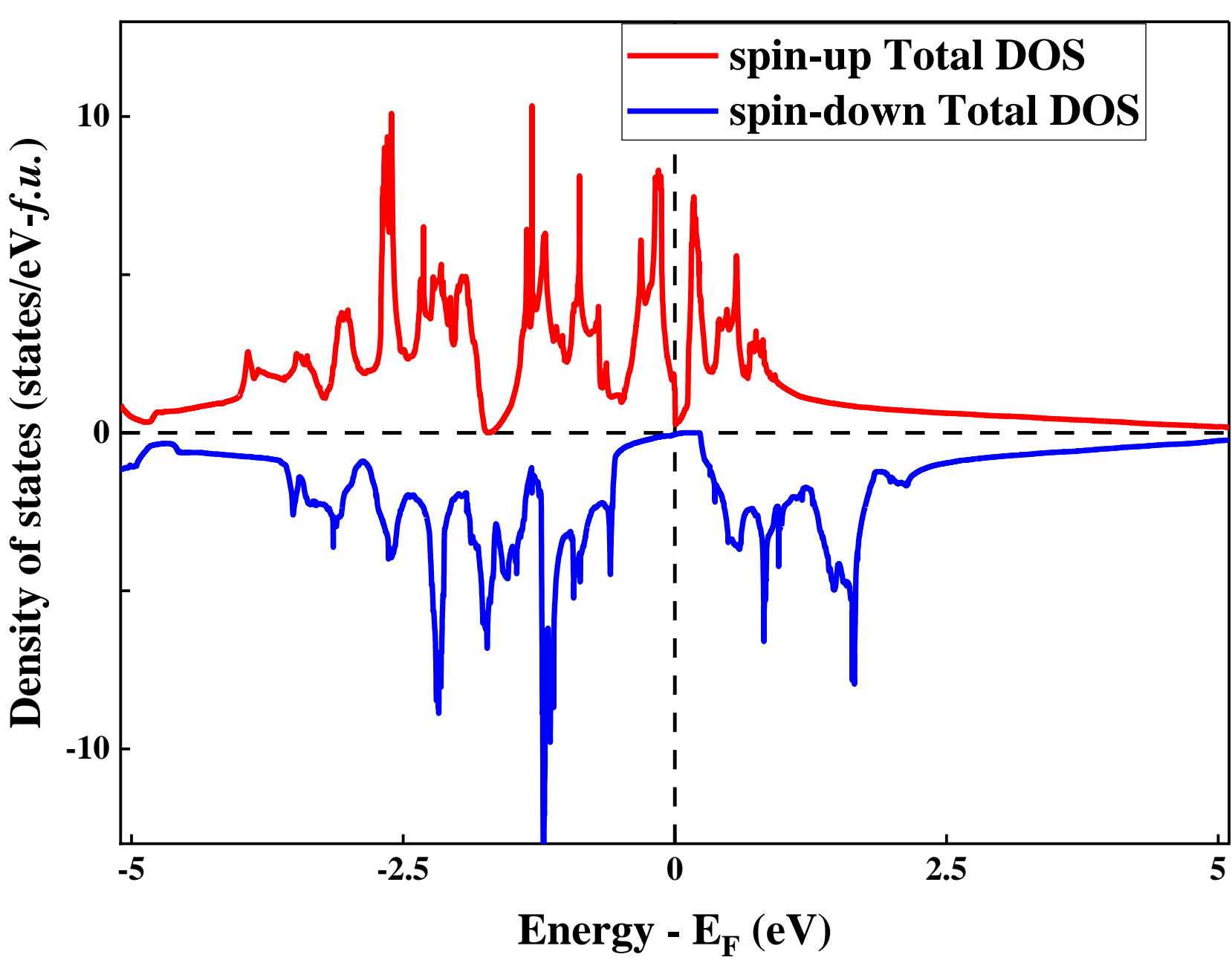

**Figure S1**: Spin-resolved density of states plot for *Y-I* ordered CoFeCrGa, displaying the SGS nature. The red and blue solid lines represent the band dispersion for spin-up and spin-down electrons.

## II. MSA for the bulk tetragonally distorted structures: convergence of MSA *w.r.t. cut-off radius*

As also discuss earlier in MS, Magnetic shape anisotropy (MSA) for lattice distorted structures is calculated as the difference in magnetic dipole-dipole interaction energy along the [100] and [001] directions, with the dipole-dipole interaction energy computed straightforwardly using Equation 3 of MS. Owing to the long-range nature of dipole-dipole interactions, the summation in Equation 3 should, in principle, run over an infinite discrete lattice for accurate determination of dipole-dipole interaction energy and, consequently MSA. However, because the interaction energy of sufficiently distant dipoles becomes negligible, a cut-off sphere (characterized by a cut-off radius) can be employed for computational simplicity, as also adopted in refs. 66–68. This approach is commonly referred to as the *direct sum method* in the literature [1].

Notably, to determine the adequate radius ($r_{cut}$) for such a cut-off sphere, the convergence of the dipole-dipole interaction energy or MSA with respect to $r_{cut}$ should be carefully examined for distorted structures. Accordingly, MSA is calculated for different rcut for one representative distorted structure (c/a = 0.8), and is plotted in Figure S2, to check out the converged $r_{cut}$. Here, it is presumed that other distorted structures would display a similar qualitative dependence on $r_{cut}$. As shown in Figure S2, it is found that MSA converges well for the $r_{cut}$ = 200 Å. In addition to MSA, the dipole-dipole interaction energies in each direction ([100] and [001]) also show a similar convergence trend. In particular, within the cut-off sphere of $r_{cut}$ = 200 Å, the dipole-dipole interaction energy for the most distant pair inside the sphere is of the order of $10^{-30}$ J, compared to a total dipole-dipole interaction energy of $10^{-25}$ J. Hence, increasing $r_{cut}$ would not significantly alter the dipolar interaction energy or resulting MSA, indicating that $r_{cut}$ = 200 Å is sufficient to calculate the dipole-dipole interaction energy and MSA. This cut-off radius is therefore adopted for all distorted structures for the calculation of MSA, together with the magnetic moments obtained from SCF calculations, as shown in Figure 3 of the manuscript.

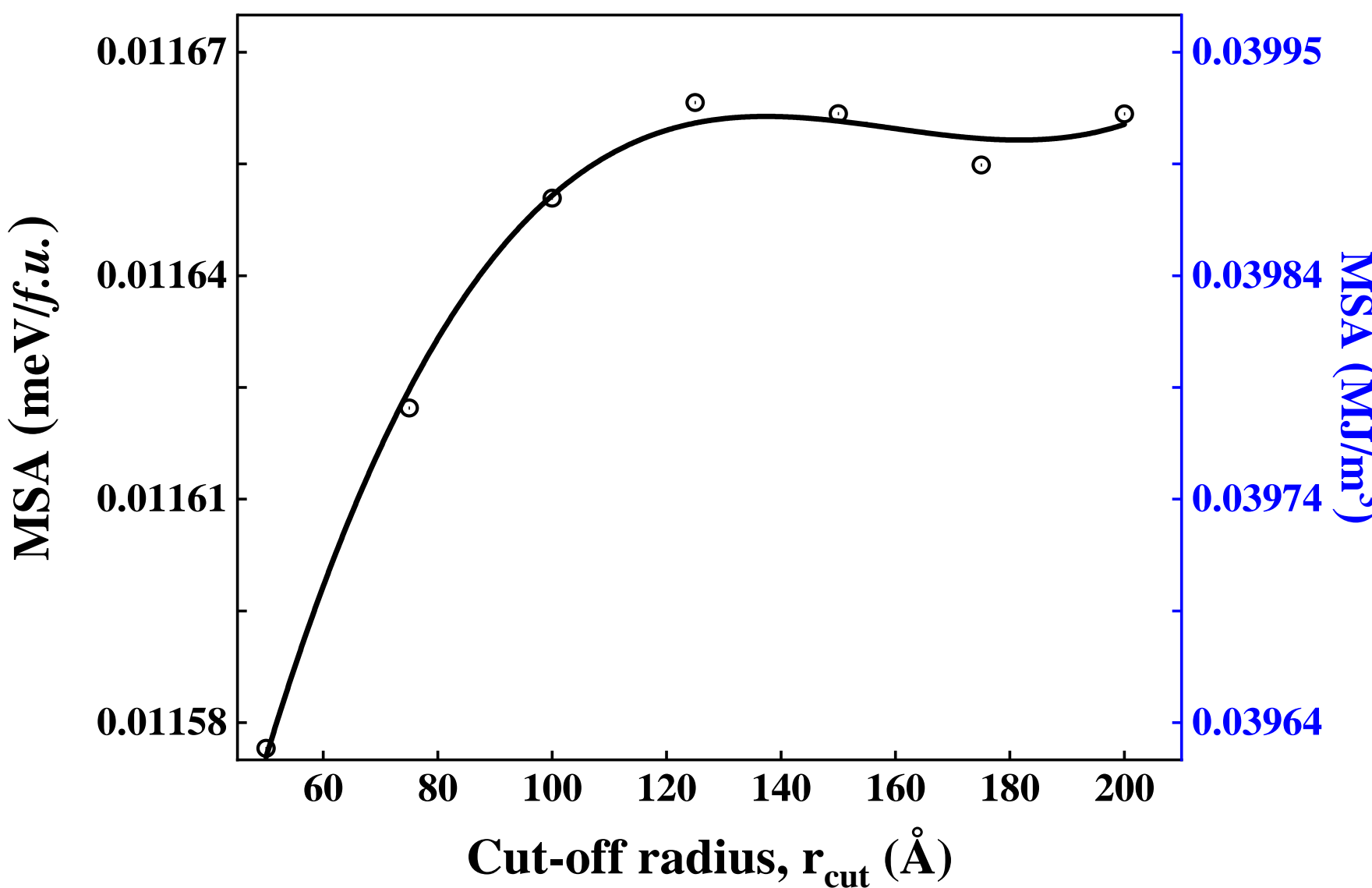

**Figure S2**: Convergence of magnetic shape anisotropy (MSA) with respect to the cut-off radius for the tetragonally distorted structure corresponding to *c/a* = 0.8.

Here, it is also important to note that the MSA is calculated using a finite-sized cut-off sphere, which may introduce small errors due to finite-size effects. The finite size effect makes it unsuitable for accurately capturing long-range interactions of magnetic ions; consequently, the direct summation method sometimes tends to overestimate the MSA. More accurate MSA can be obtained using advanced techniques, such as the Ewald summation method or the particle-particle-mesh (P3M) method, which better account for long-range interactions. However, since the MCA is significantly larger than the corresponding MSA, and the MSA is already slightly overestimated, any further improvement in MSA would not affect the overall conclusions regarding the total magnetic anisotropy.